\documentclass[prl,twocolumn,showpacs]{revtex4}
\def\address{\affiliation}
\usepackage[dvips]{graphicx}
\begin{document}

\title{
Large Transport Properties of Bi$_2$Sr$_{2-x}$La$_x$CaCu$_2$O$_{8+\delta}$ Single Crystals Grown by a Floating-Zone Method}

\author{
Takenori Fujii$^{1,2}$\footnote{E-mail address:fujii@htsc.sci.waseda.ac.jp} and Ichiro Terasaki$^{1,2}$.
}

\address{
$^1$Department of Applied Physics, Waseda University, Tokyo 169-8555, Japan.\\
$^2$PREST, Japan Science and Technology Corporation, Kawaguchi 332-0012, JAPAN.\\}%

\begin{abstract}
A parent insulator of the Bi-based high-$T_c$ superconductor Bi$_2$Sr$_{2-x}$La$_x$CaCu$_2$O$_{8+\delta}$ was successfully grown by a traveling solvent floating zone method. The in-plane resistivity $\rho_{ab}$ shows metallic behavior ($d\rho_{ab}/dT >$ 0) near room temperature, although the magnitude of the resistivity is far above the Mott limit for metallic conduction. The Hall mobility of the parent insulator does not so much differ from that of an optimally doped sample. These results suggest that the system is not a simple insulator with a well-defined gap, and the hole in the parent insulator is essentially itinerant, as soon as it is doped. 
\end{abstract}

\date{\today}

\pacs{74. 25. Fy, 74. 25. Dw, 74. 72. Hs}

\maketitle

\section{Introduction}
It is well known that the high $T_c$ superconductivity occurs with doping carriers into the antiferromagnetic (AF) insulator. The AF insulator is quite different from a conventional insulator, and shows various peculiar features. One of the most peculiar features is that in the lightly doped La$_{2-x}$Sr$_x$CuO$_4$ (LSCO) metallic behavior of the in-plane resistivity $\rho_{ab}$ ($d\rho_{ab}/dT >$ 0) near room temperature was observed, even though the magnitude of the in-plane resistivity is far above the Mott limit~\cite{ando}. This metallic behavior is considered to be responsible for the appearance of band dispersion within the Mott-Hubbard gap seen in the angle-resolved photoemission spectroscopy (ARPES) measurement~\cite{teppei}. The electric states of the doped parent insulator, on the other hand, seem to be quite different between LSCO and the Bi-based cuprates. The charge-stripe order is stabilized in LSCO, whereas it is much weaker in the Bi-based cuprates. LSCO has been most frequently used for systematic studies of the heavily underdoped region of the cuprates, because a heavily underdoped sample is easy to synthesize. By contrast, there are only a few reports on the lightly doped Bi$_2$Sr$_2$CaCu$_2$O$_{8+\delta}$ ~\cite{terra, kend}, because it is difficult to grow high-quality single crystals of lightly doped region. We have previously grown the parent insulator of Bi$_2$Sr$_2$CaCu$_2$O$_{8+\delta}$ by flux method, where the trivalent lanthanide, such as Dy and Er, is substituted for divalent Ca~\cite{terra}. Here, we have successfully grown the parent insulator of Bi$_2$Sr$_{2-x}$La$_x$CaCu$_2$O$_{8+\delta}$ by a traveling solvent floating zone (TSFZ) method. 

\section{Experimental}
Two crystals of 40\% La doped Bi$_2$Sr$_2$CaCu$_2$O$_{8+\delta}$ were used in this study. One is a crystal in which all La atoms were substituted for Sr site (Bi$_2$(Sr,La)$_2$CaCu$_2$O$_{8+\delta}$), and the other is a crystal in which La atoms were substituted for Sr and Ca sites (Bi$_2$(Sr,Ca,La)$_3$Cu$_2$O$_{8+\delta}$). They were grown by TSFZ method with a growth velocity of 0.5mm/h in a mixed gas flow of O$_2$ (20\%) and Ar (80\%). The starting composition of the feed rods were Bi$_{2.1}$Sr$_{1.1}$La$_{0.8}$Ca$_{1.0}$Cu$_{2.0}$O$_{8+\delta}$ for Bi$_2$(Sr,La)$_2$CaCu$_2$O$_{8+\delta}$ and Bi$_{2.1}$Sr$_{1.5}$La$_{0.8}$Ca$_{1.0}$Cu$_{2.0}$O$_{8+\delta}$ for Bi$_2$(Sr,Ca,La)$_3$Cu$_2$O$_{8+\delta}$. Very large single crystals with a size of 2 $\times$ 4 $\times$ 0.1 mm$^3$ were successfully grown. By energy dispersive x-ray analysis (EDX), the actual compositions were determined to be Bi$_{2.01}$Sr$_{1.11}$La$_{0.84}$Ca$_{1.00}$Cu$_{2.00}$O$_{8+\delta}$ for Bi$_2$(Sr,La)$_2$CaCu$_2$O$_{8+\delta}$, and Bi$_{2.04}$Sr$_{1.31}$La$_{0.86}$Ca$_{0.76}$Cu$_{2.00}$O$_{8+\delta}$ for Bi$_2$(Sr,Ca,La)$_3$Cu$_2$O$_{8+\delta}$. In the case of Bi$_2$(Sr,Ca,La)$_3$Cu$_2$O$_{8+\delta}$, excess La replaces the Ca site. Figure 1 shows the XRD pattern for the as-grown Bi$_2$(Sr,La)$_2$CaCu$_2$O$_{8+\delta}$ single crystal. All the peaks were indexed as (0 0 $l$) peaks of the Bi-2212 phase with no trace of impurities. The $c$-axis length was evaluated to be 30.18 \AA~ for Bi$_2$(Sr,La)$_2$CaCu$_2$O$_{8+\delta}$, and 30.38 \AA~ for Bi$_2$(Sr,Ca,La)$_3$Cu$_2$O$_{8+\delta}$, which is smaller than that of Bi$_2$Sr$_2$CaCu$_2$O$_{8+\delta}$ (30.86 \AA). This is due to the larger ion radius of the Sr atom (1.13 \AA) than that of the La (1.05 \AA) and Ca (1.00 \AA) atoms. The in-plane resistivity of Bi$_2$(Sr,La)$_2$CaCu$_2$O$_{8+\delta}$ was measured using standard four-probe method, whereas the in-plane resistivity of the highly insulating sample Bi$_2$(Sr,Ca,La)$_3$Cu$_2$O$_{8+\delta}$ was measured using a two-probe method with constant voltage (where the sample resistance is much higher than the contact resistance). The voltage for both samples was confirmed to vary linearly with the current in the measured range from 4.2 to 300 K. The thermopower was measured using a steady-state technique from 4.2 to 300 K. The Hall coefficient measurement was done by sweeping the magnetic field which was applied along the c axis (c // H) from -7 to 7 T at fixed temperatures. The Hall resistance was confirmed to be linear in magnetic field. The crystals were annealed in the O$_2$ or Ar atmosphere at 600 $^{\circ}$C for 10 h to change the oxygen contents as is similar to Bi$_2$Sr$_2$CaCu$_2$O$_{8+\delta}$~\cite{wata}. 

\begin{figure}
 \includegraphics[width=7cm,clip]{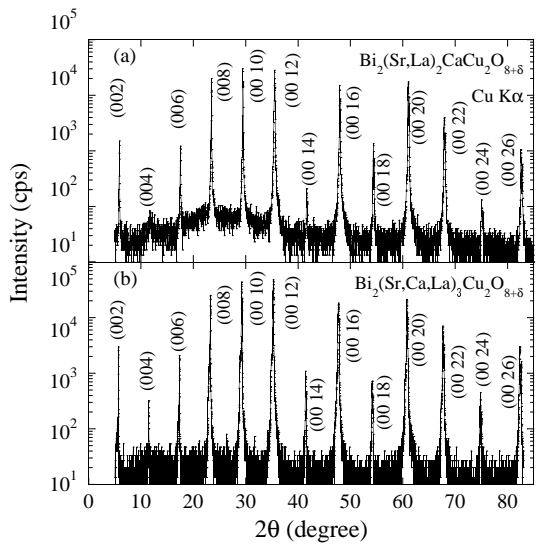}
 \caption{
 X-ray diffraction pattern for (a) as-grown Bi$_2$(Sr,La)$_2$CaCu$_2$O$_{8+\delta}$ (b) as-grown Bi$_2$(Sr,Ca,La)$_3$Cu$_2$O$_{8+\delta}$ single crystals.
 }
\end{figure}

\section{Results and discussion}
Figure 2 shows the temperature dependence of the in-plane resistivities for Bi$_2$(Sr,La)$_2$CaCu$_2$O$_{8+\delta}$ (closed circles) and Bi$_2$(Sr,Ca,La)$_3$Cu$_2$O$_{8+\delta}$ (open circles). Bi$_2$(Sr,Ca,La)$_3$Cu$_2$O$_{8+\delta}$ shows insulating behavior ($d\rho_{ab}/dT$ $<$ 0) in the measured temperature range. The magnitude of the in-plane resistivity at room temperature is about 30 m$\Omega$cm for an as-grown sample. It increases to 120 m$\Omega$cm by annealing in Ar, which suggests that the carrier concentration decreases with decreasing the oxygen content. Bi$_2$(Sr,La)$_2$CaCu$_2$O$_{8+\delta}$ shows metallic behavior ($d\rho_{ab}/dT$ $>$ 0) above 150 K, and the magnitude of the in-plane resistivity is about 5 m$\Omega$cm at room temperature. In the inset of Fig. 2, the in-plane resistivity of Bi$_2$(Sr,La)$_2$CaCu$_2$O$_{8+\delta}$ is plotted in linear scale to see the metallic behavior clearly. Although observed metallic behavior seems to suggest the Boltzmann transport, the magnitude of the in-plane resistivity is far above the Mott limit for metallic conduction, which is quite different from a conventional insulator (usually called as a ``bad metal''~\cite{emery}). 
\begin{figure}
 \includegraphics[width=7cm,clip]{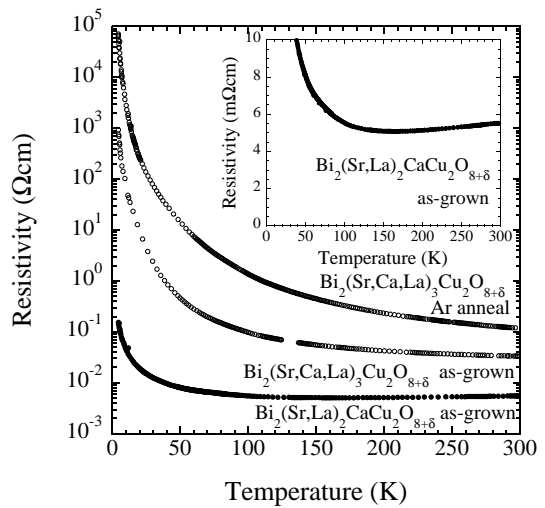}
 \caption{
 Temperature dependence of the in-plane resistivities $\rho_{ab}$ for Bi$_2$(Sr,La)$_2$CaCu$_2$O$_{8+\delta}$ (closed circles) and Bi$_2$(Sr,Ca,La)$_3$Cu$_2$O$_{8+\delta}$ (open circles) with various annealing conditions. Inset: $\rho_{ab}$ of Bi$_2$(Sr,La)$_2$CaCu$_2$O$_{8+\delta}$ is plotted in linear scale to emphasize metallic behavior.
 }
\end{figure}

Figure 3 shows the temperature dependence of the in-plane thermopower for Bi$_2$(Sr,La)$_2$CaCu$_2$O$_{8+\delta}$ (closed circles) and Bi$_2$(Sr,Ca,La)$_3$Cu$_2$O$_{8+\delta}$ (open circles). For both samples, the thermopower decreases with decreasing temperature, which is consistent with our previous results of Bi$_2$Sr$_2$RCu$_2$O$_{8+\delta}$ (R=Dy, Er)~\cite{terra2}. The magnitude of the thermopower increases by Ar annealing and decreases by O$_2$ annealing, which means that the thermopower decreases with increasing oxygen contents. Since the room temperature thermopower is considered to be a universal measure of the doping level~\cite{obert}, this result indicates that the carrier doping is caused by the excess oxygen. Note that the room-temperature thermopower and the magnitude of the in-plane resistivity of Bi$_2$(Sr,La)$_2$CaCu$_2$O$_{8+\delta}$ are smaller than those of Bi$_2$(Sr,Ca,La)$_3$Cu$_2$O$_{8+\delta}$. Although these two compounds contain approximately the same amount of La ($x$ $\sim$ 0.8), the carrier concentration is different. For example, the carrier concentration per Cu $p$ are estimated to be 0.02 for as-grown Bi$_2$(Sr,Ca,La)$_3$Cu$_2$O$_{8+\delta}$, and 0.04 for as-grown Bi$_2$(Sr,La)$_2$CaCu$_2$O$_{8+\delta}$. One explanation for this is that the substitution of La for Ca site causes a disorder into the CuO$_2$ plane to make the doped carriers localized. Another explanation is that the amount of the excess oxygen is different. 

\begin{figure}
 \includegraphics[width=7cm,clip]{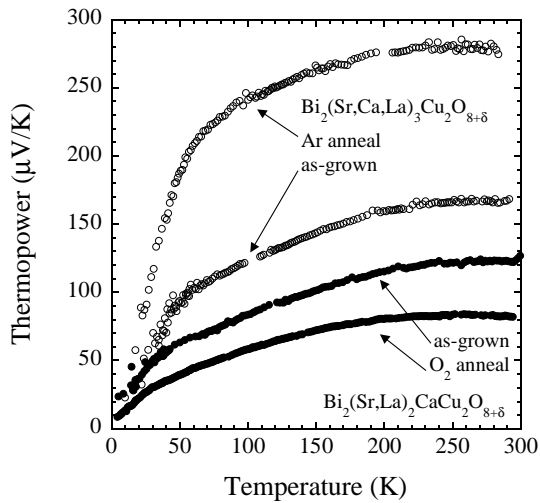}
 \caption{
 Temperature dependence of the thermopower for Bi$_2$(Sr,La)$_2$CaCu$_2$O$_{8+\delta}$ (closed circles) and Bi$_2$(Sr,Ca,La)$_3$Cu$_2$O$_{8+\delta}$ (open circles) with various annealing conditions.
 }
\end{figure}

\begin{figure}
 \includegraphics[width=7cm,clip]{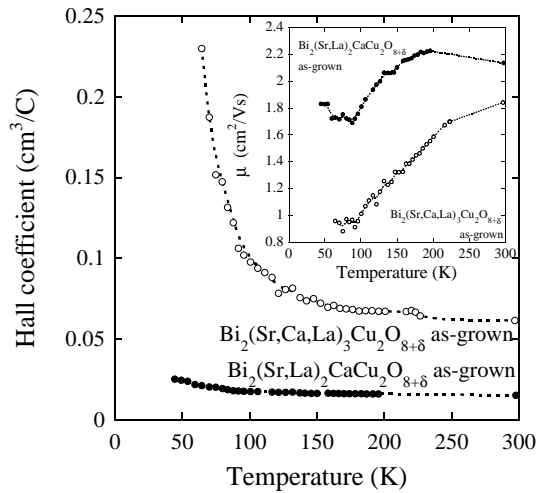}
 \caption{
 Temperature dependence of the Hall coefficient for as-grown Bi$_2$(Sr,La)$_2$CaCu$_2$O$_{8+\delta}$ (closed circles) and as-grown Bi$_2$(Sr,Ca,La)$_3$Cu$_2$O$_{8+\delta}$ (open circles). Inset: Hall mobility $\mu \equiv R_H/\rho_{ab}$ plotted as a function of temperature.
 }
\end{figure}

Figure 4 shows the temperature dependence of the Hall coefficient R$_H$ for as-grown Bi$_2$(Sr,La)$_2$CaCu$_2$O$_{8+\delta}$ (closed circles) and as-grown Bi$_2$(Sr,Ca,La)$_3$Cu$_2$O$_{8+\delta}$ (open circles). The Hall coefficients monotonically increase with decreasing temperature for both samples. The room temperature carrier concentration is estimated from the relation $n = (eR_H)^{-1}$ to be 4 $\times$ 10$^{20}$ cm$^{-3}$ ($p \sim$ 0.04) for Bi$_2$(Sr,La)$_2$CaCu$_2$O$_{8+\delta}$, and 1 $\times$ 10$^{20}$ cm$^{-3}$ ($p \sim$ 0.01) for Bi$_2$(Sr,Ca,La)$_3$Cu$_2$O$_{8+\delta}$, which agree well with the carrier concentration estimated from the room temperature thermopower. The Hall mobility calculated using the Hall coefficient and the resistivity as $\mu \equiv R_H/\rho_{ab}$ is shown in the inset of Fig. 4. It should be emphasized that the mobility is relatively large; the magnitude of which (2.2 cm$^2$/Vs for Bi$_2$(Sr,La)$_2$CaCu$_2$O$_{8+\delta}$ and 1.8 cm$^2$/Vs for Bi$_2$(Sr,Ca,La)$_3$Cu$_2$O$_{8+\delta}$) is only 5 times smaller than that of Bi$_2$Sr$_2$CaCu$_2$O$_{8+\delta}$ (10 cm$^2$/Vs)~\cite{maeda}, although the magnitude of the in-plane resistivity of Bi$_2$(Sr,La,Ca)$_3$Cu$_2$O$_{8+\delta}$ is 100 times larger than that of Bi$_2$Sr$_2$CaCu$_2$O$_{8+\delta}$. In our previous work of the parent insulator Bi$_2$Sr$_2$RCu$_2$O$_{8+\delta}$ (R = Dy, Er), the mobility was too small to measure the Hall coefficient precisely~\cite{syuuron}. A similar large mobility at 300 K was reported in the lightly doped LSCO, where only 1\% of hole doping causes the metallic behavior~\cite{ando}. 

In the case of a conventional semiconductor, the mobility does not decrease with decreasing carrier concentration. It is rather possible that the mobility becomes large with decreasing carrier concentration because of the suppression of the electron-electron scattering. While, in the case of the doped antiferromagnetic insulator, such as high-$T_c$ superconductor, the hole is replaced by the spin in the next site whenever it moves. Then, if the hole moves independently, there is the energy loss of the order of the antiferromagnetic correlation energy and the mean free pass becomes as small as the lattice constant. The fact that the mobility of the parent insulator does not so much differ from that of an optimally doped sample does not agree with above simple consideration. It strongly suggests that the doped holes form a charge order, such as charge stripe or phase separation, and move in a group. Thus, the doped hole in the parent insulator is essentially itinerant and shows metallic behavior.

\section{Conclusion}

In conclusion, we have successfully grown large single crystals of the parent insulator of Bi$_2$Sr$_2$CaCu$_2$O$_{8+\delta}$ by substituting La for Sr. With using these crystals, we have measured the in-plane resistivity , thermopower and Hall coefficient along the CuO$_2$ plane. the in-plane resistivity of the most conducting sample shows metallic behavior above 150 K. Although the magnitude of the in-plane resistivity of the parent insulator is about 100 times larger than that of optimum doped Bi$_2$Sr$_2$CaCu$_2$O$_{8+\delta}$, the mobility is not so much different from that of the superconducting compounds. These results suggest that the system is not a conventional insulator with a well-defined gap, and the hole in the parent insulator is essentially itinerant, as soon as it is doped. Our single crystal shows relatively large mobility compared to Ca-site-substituted single crystals, which indicates a smaller disorder in CuO$_2$ plane.

\end{document}